\documentclass[prl,twocolumn,10pt]{revtex4-2}
\usepackage{bm}
\usepackage{epsfig}
\usepackage{bbold}
\usepackage{graphicx}
\usepackage{amsmath}
\usepackage{amssymb}
\usepackage{amsbsy}
\usepackage{color}
\usepackage{subfigure}
\usepackage{nicefrac}
\usepackage{slashed}
\usepackage{afterpage}
\usepackage{psfrag}
\usepackage{hyperref}
\usepackage[capitalise]{cleveref}
\usepackage{dsfont}

\newcommand{\beq}{\begin{equation}}
\newcommand{\eeq}{\end{equation}}
\newcommand{\bea}{\begin{eqnarray}}
\newcommand{\eea}{\end{eqnarray}}

\DeclareMathOperator{\GeV}{GeV}

\DeclareMathOperator{\keV}{keV}

\allowdisplaybreaks

\begin{document}

\title{Do direct detection experiments constrain axionlike particles coupled to electrons?}

\author{Ricardo Z.~Ferreira}
\email{rzambujal@ifae.es}
\affiliation{Institut de F\'isica d'Altes Energies (IFAE) and Barcelona Institute of Science and Technology (BIST),
		Campus UAB, 08193 Bellaterra, Barcelona, Spain}

\author{M.C.~David~Marsh and Eike M\"uller}
\email{david.marsh@fysik.su.se}
\email{eike.muller@fysik.su.se}
\affiliation{The Oskar Klein Centre for Cosmoparticle Physics,
Department of Physics,
Stockholm University, AlbaNova, 10691 Stockholm, Sweden
\looseness=-1}

\date{\today}

\begin{abstract}
Several laboratory experiments have published limits on axionlike particles (ALPs) with feeble couplings to electrons and masses in the keV--MeV range, under the assumption that such ALPs comprise the dark matter. We note that ALPs  decay radiatively into photons, and show that for a large subset of the parameter space ostensibly probed by these experiments, the lifetime of the ALPs is shorter than the age of the universe. Such ALPs cannot consistently make up the dark matter, which significantly affects the interpretation of published limits from GERDA, Edelweiss-III, SuperCDMS and Majorana. Moreover, constraints from X-ray and gamma-ray astronomy exclude an even wider range of the ALP-electron coupling, and supersede all current laboratory limits on dark matter ALPs in the 6 keV to 1 MeV mass range. These conclusions are rather model-independent, and can only be avoided at the expense of significant fine-tuning in theories where the ALP has additional couplings to other particles. 
\end{abstract}

\maketitle

\section{Introduction}
The nature of dark matter is one of the most central, open questions of contemporary fundamental physics.
Axionlike particles (ALPs) are pseudo-scalar remnants of broken symmetries, and appear frequently in theories beyond the Standard Model. ALPs can comprise the dark matter for a broad range of masses, and over the past few years, several dark matter `direct detection' experiments (as well as experiments designed to observe neutrinoless double-$\beta$ decays), have published limits on `heavy' ALP dark matter with masses in the keV--MeV range \cite{Panda:2017,Majorana:2016hop,XMASS:2018pvs,EDELWEISS:2018tde,SuperCDMS:2019jxx,XENON:2020rca,GERDA:2020emj, LUX}. 
An important assumption of these experimental searches is the existence of a galactic density of dark matter particles that can interact with the detector.

\begin{figure}[b]
    \centering
    \includegraphics[width=.35\textwidth]{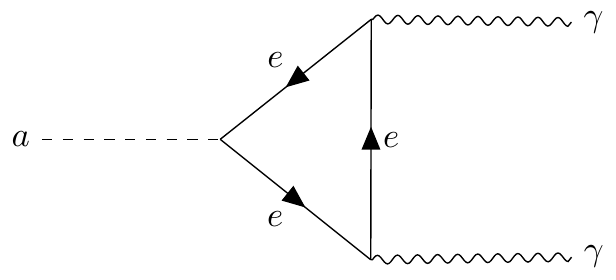}
    \caption{The decay of an ALP through an off-shell electron loop (both loop orientations contribute).}
    \label{fig:triangleDiagram}
\end{figure}

In this paper, we point out that heavy ALPs coupled only to electrons, or with generic couplings to other particles, decay quickly into photons. Short-lived ALPs cannot comprise the dark matter, and direct detection limits on the couplings of such particles are not meaningful. Moreover, we point out that stringent limits from `indirect detection' via X-ray and gamma-ray signals exclude ALPs coupled to electrons over most of the parameter space that current and future generations of direct detection experiments are sensitive to. We end by showing that these arguments are generically robust also in ALP theories with additional couplings to other particles, but point out fine-tuned slivers of the extended parameter space in which the stability issues can be avoided.

\section{Axionlike particles decay into photons}
Experiments using liquid xenon \cite{Panda:2017, XMASS:2018pvs, XENON:2020rca, LUX} 
and germanium silicon \cite{Majorana:2016hop, SuperCDMS:2019jxx, EDELWEISS:2018tde, GERDA:2020emj} can be sensitive to the absorption of galactic ALPs in the detector materials through a process similar to the photo-electric effect. This process depends  on the (dimensionless) coupling of ALPs to electrons, $g_{ae}$, which is defined by the Lagrangian density
\begin{equation}
    {\cal L} \supset
    \frac{1}{2} \partial_\mu a \partial^\mu a - \frac{1}{2} m_a^2 a^2 + \frac{g_{ae}}{2 m_e} \partial_\mu a \,  \bar \psi_e \gamma^\mu \gamma^5 \psi_e \, ,
    \label{eq:L1}
\end{equation}
where $a$ denotes the ALP (of mass $m_a$), and $\psi_e$ denotes the Dirac field for the electrons (of mass $m_e$). For tree-level processes such as ALP-absorption, the derivative-interaction term can equivalently be written in the pseudo-scalar form, $ -i g_{ae} a  \bar \psi_e \gamma^5 \psi_e$, which is commonly used in the literature. However, quantum mechanically these interactions are inequivalent and differ by anomalous couplings (in the massless limit), e.g.~by an additional coupling to photons \cite{Bauer:2017ris,Caputo:2021rux} (see also \cite{Takahashi:2020bpq} for a related discussion in the context of the XENON1T excess \cite{XENON:2020rca}). In this paper, we use the derivative form in Eq. \eqref{eq:L1}, which makes the underlying ALP shift symmetry manifest. This choice is also conservative, as we make clear below.

A viable dark matter candidate must have a lifetime that is longer than the age of the universe.  
For $ x \equiv m_a/(2m_e) < 1 $, ALPs  cannot decay into electrons due to energy conservation, and are stable at tree-level.
However, ALPs described by the Lagrangian of Eq.~\eqref{eq:L1} inevitably decay into two photons through the well-known one-loop process involving the triangle diagram with off-shell electrons shown in \cref{fig:triangleDiagram}. \begin{figure*}
 	\centering
	\includegraphics[width=0.9\linewidth]{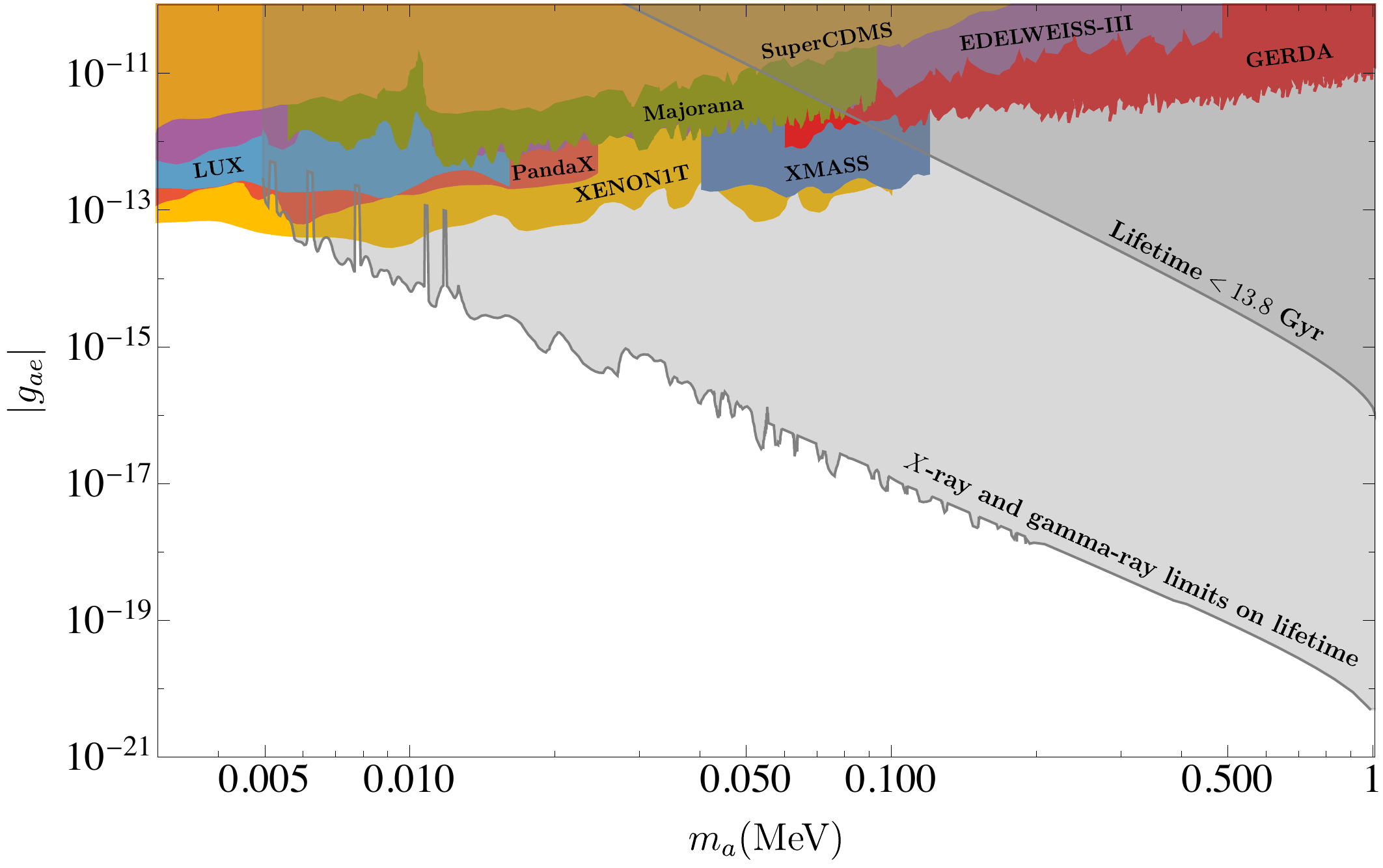}
	\caption{Limits on $g_{ae}$ from direct detection searches for dark matter axionlike particles. Parameter region with an ALP-lifetime shorter than the age of the universe (dark grey), and the region excluded by X-ray and gamma-ray searches (light grey).}
	\label{fig:Constraints on gae for axion dark matter}
\end{figure*}
The  $a \to \gamma \gamma$ decay rate is given by
\begin{eqnarray}
	\Gamma_{a \to \gamma\gamma} &=& \frac{m_a^3\, (g^{\text{1-loop}}_{a\gamma \gamma})^{2}}{64 \pi} \, ,
	\label{eq:gamma}
	\end{eqnarray}
where the effective ALP-photon coupling for $ x<1 $ is given by \cite{Pospelov:2008jk,Bauer:2017ris}
\begin{eqnarray}
g^{\text{1-loop}}_{a\gamma \gamma}&= &\frac{ \alpha g_{ae}}{m_e \pi} \left(1 - x^{-2} \arcsin^2(x)\right) \, .
\label{eq:gag}
\end{eqnarray}
Here $\alpha$ is the fine-structure constant. We note that Eq. \eqref{eq:gag} gives the coupling for the decay process, and is in general different from the one-loop effective coupling relevant for other processes like the Primakoff effect \cite{Ferreira:2022}.

An important observation is that, for ALP masses in the keV--MeV range, the virtual electrons are not far off shell, and the resulting decay rate can be large. Expanding to leading order in $ x \ll 1 $, the loop-induced coupling is given by
\begin{align} \label{eq: gag loop large x}
    g^{\text{1-loop}}_{a\gamma \gamma}
    &\approx - \frac{\alpha g_{ae} m_a^2}{12\pi m_e^3}\\
    &\approx - 1.5 \cdot 10^{-14} \GeV^{-1} \left( \frac{g_{ae}}{10^{-12}}\right)\left(\frac{m_a}{100 \keV} \right)^2\, ,\nonumber
\end{align}
and the resulting lifetime $ \tau_{a \to \gamma\gamma} = \Gamma_{a \to \gamma\gamma}^{-1} $ is given by
\begin{equation}\label{eq:gammaexp}
	\frac{\tau_{a \to \gamma\gamma}}{13.8\, \text{Gyr}} \approx 
	\left(\frac{1.2 \cdot 10^{-12}}{g_{ae}} \right)^2 \left( \frac{100\, \text{keV}}{m_{a}}\right)^7 \, .
\end{equation}
The lifetime depends strongly on the ALP mass, and the radiative decay 
cannot be ignored for heavy ALPs. 
Note that, if we had defined $g_{ae}$ through the pseudo-scalar coupling, there would be an additional, $m_a$-independent contribution ($\sim\alpha g_{ae}/(m_e \pi)$) to the ALP-photon coupling \cite{Bauer:2017ris,Caputo:2021rux}, which would further reduce the ALP life-time. For this reason, our choice of interaction term in Eq.~\eqref{eq:L1} is conservative. 

In \cref{fig:Constraints on gae for axion dark matter} we show (in dark grey) the region in which $ \tau_{a \to \gamma\gamma} $ is shorter than the age of the universe. The experimental dark matter bounds that extend into this region are not self-consistent within the theory given by Eq.~\eqref{eq:L1}.

Most severely affected by this issue is the GERDA experiment that ostensibly probes dark matter ALPs coupled to electrons with masses from 60 keV to 1 MeV \cite{GERDA:2020emj} (shown in dark red in \cref{fig:Constraints on gae for axion dark matter}).
In the lower mass region of $m_a = 60$--$88$ keV,
GERDA excludes only a small triangle in the parameter space when the lifetime condition $\tau_{a \to \gamma\gamma}>13.8$ Gyr is imposed.  
For the remaining 97\% of the mass range (88 keV -- 1 MeV),
no self-consistent bound on dark matter ALPs described by  Eq.~\eqref{eq:L1} can be inferred from the GERDA search.

Other experiments that have published limits that are significantly affected by this stability issue include
SuperCDMS \cite{SuperCDMS:2019jxx}, Majorana \cite{Majorana:2016hop} and EDELWEISS-III \cite{EDELWEISS:2018tde}. To a lesser extent, XENON1T \cite{XENON:2020rca} and XMASS \cite{XMASS:2018pvs} are also affected, but these experiments still constrain some interval of $g_{ae}$ for any mass within the published ranges.

More severe constraints on the decay $a \to \gamma \gamma$ in the keV--MeV mass range come from searches for the corresponding X-ray and gamma-ray signals in the Milky Way and M31 galaxy using the data obtained from the \textit{NuStar}, \textit{XMM-Newton} and \textit{INTEGRAL} telescopes \cite{Ng:2019gch,Laha:2020ivk, Foster:2021ngm}. No such signal has been observed, which leads to stringent limits on $g_{ae}$ (cf.~the light grey region in \cref{fig:Constraints on gae for axion dark matter}). This indirect limit excludes the parameter region probed by all current  experimental searches for $ m_a > 6 $ keV.

\section{ALP theories with additional couplings}

The presence of tree-level couplings to other particles than the electron, e.g.~photons or other electrically charged fermions, results in further contributions to the decay rate, on top of the one proportional to $g_{ae}$. Such contributions typically add to the decay rate and consequently make the ALP lifetime conditions more severe. However, at the expense of significantly fine-tuning the additional couplings, it is possible to prolong the ALP lifetime and conceive a setup where the experimental searches are self-consistent. In this section we show, via two examples, how delicate such constructions need to be.

First, consider a theory that, in addition to the Lagrangian in \cref{eq:L1}, includes a tree-level coupling of ALPs to photons of the form
\begin{eqnarray}
    {\cal L} \supset 
    \frac{g^{\text{tree}}_{a \gamma \gamma}}{4}
a \, F_{\mu \nu} \tilde{F}^{\mu \nu}\, ,
\end{eqnarray}
where $F$ is the electromagnetic field strength tensor and $\tilde{F}$ its dual. The decay rate is now given by Eq.~\eqref{eq:gamma}, but with the one-loop coupling replaced by the sum $g_{a\gamma\gamma}^{\rm tree} + g^{\text{1-loop}}_{a\gamma \gamma}$. In order to prolong the ALP lifetime, the tree-level coupling has to cancel against the one-loop contribution from electrons 
such that
\begin{eqnarray}
    g^{\text{tree}}_{a \gamma \gamma} 
    = -g^{\text{1-loop}}_{a \gamma \gamma}
    \simeq \frac{\alpha g_{ae} m_a^2}{12\pi m_e^3}\, ,
    \label{eq:finetune1}
\end{eqnarray}
where the last step applies to the $x\ll1$ limit.
This equality
has to be satisfied to a high accuracy. For example, in order to realise an ALP dark matter model with $m_a = 1$ MeV, an electron-coupling within reach of GERDA ($g_{ae} \gtrsim 10^{-11}$), and a lifetime longer than the age of the universe, the equality~\eqref{eq:finetune1} needs to hold to a precision of at least one part in $10^5$. In order to avoid the astrophysical gamma-ray constraints, the agreement would have to be better than one part in $10^9$. Note also that, because the tuning depends on the ALP mass, each point in the ($m_a,g_{ae}$) parameter space would correspond to a different tree-level coupling to photons. 

Second, consider a theory that, in addition to the Lagrangian in \cref{eq:L1}, includes a tree-level coupling of ALPs to a heavy lepton ($\psi_i$, with $ i = \mu,\, \tau$) of the form
\begin{eqnarray}
   {\cal L} \supset \frac{g_{ai}}{2 m_i} \partial_\mu a \,  \bar \psi_i \gamma^\mu \gamma^5 \psi_i \,.
\end{eqnarray}
The loop-induced coupling to photons then receives an additional contribution
\begin{equation}
\begin{split}
g^{\text{1-loop}}_{a\gamma \gamma} = \frac{ \alpha }{\pi} \, \Big[ \frac{g_{ae}}{m_e}\left(1 - x^{-2} \arcsin^2(x)\right) \\+  \frac{ g_{a i}}{m_i}\left(1 - y^{-2} \arcsin^2(y)\right) \,\Big]\, ,
\label{eq:gag2}
\end{split}
\end{equation}
where $y=m_a/(2m_i)$ and $m_i$ is the mass of the heavy lepton $\psi_i$.  We expand Eq.~\eqref{eq:gag2} to leading order in $x,y\ll 1$, and require that the two loop contributions cancel each other in order to prolong the ALP lifetime. This results in a mass-enhanced coupling to the heavy lepton:
\beq 
g_{a i} \simeq -\left(\frac{m_i}{m_e} \right)^3 g_{ae} \, .
\label{eq:finetune2}
\eeq
Just as discussed in the first fine-tuned example above, this relation needs to hold with high accuracy in order to make ALPs long-lived enough to be the dark matter, and to avoid the astrophysical constraints.

The experiments shown in Fig.~\ref{fig:Constraints on gae for axion dark matter} that probe the ALP-electron coupling in the keV--MeV mass range have sensitivities limited to $|g_{ae}|\gtrsim 10^{-13}$.
For ALPs coupled to electrons and muons, Eq.~\eqref{eq:finetune2} requires that $|g_{a\mu}| >0.9\cdot 10^{-6}$. However, such a large ALP-muon coupling is ruled out by the cooling rate of the supernova SN1987A (giving $|g_{a\mu}| <0.9\cdot 10^{-8}$) \cite{Caputo:2021rux}. Thus, it is not possible to circumvent the $a\to \gamma \gamma$ decay by  fine-tuning electron and muon loops. 

ALPs coupled to electrons with $|g_{ae}|\gtrsim 10^{-13}$ require an ALP-tau coupling of  $|g_{a\tau}|\gtrsim 4\cdot 10^{-3}$ in order to satisfy Eq.~\eqref{eq:finetune2}. While this coupling is very large, it is to our knowledge not presently constrained in this mass range. However, the required hierarchy between the ALP-electron and ALP-tau couplings is unstable at the quantum level: the large $g_{a\tau}$ induces a running of $g_{ae}$ via the renormalisation group equations that quickly make $g_{ae}$ exceed $10^{-13}$. This can be verified using the results  of   \cite{DEramo:2018vss}, and implies that equation \eqref{eq:finetune2} must hold as an accidental relation at low energies.

\section{Outlook}
The main result of this paper is that the radiative $a\to \gamma \gamma$ decay  is of critical importance for ALP dark matter searches in the keV--MeV mass range. For a large fraction of the relevant parameter space, an ALP only coupled to electrons (or with generic couplings to other fields)  is unstable on cosmological scales, and cannot comprise the dark matter. This result has significant implications for the interpretation of published results from direct detection experiments, and is of strategic importance when planning future experimental programs.

Our results imply that a hypothetical \emph{detection} of heavy, dark matter ALPs coupled to electrons could point to delicate models in which multiple couplings are fine-tuned to cancel the $a\to \gamma \gamma$ decay rate. This would be a very informative and surprising insight into the ALP sector which is typically expected to be free of tunings.

In addition to searching for galactic dark matter ALPs, direct detection experiments can probe ALPs produced in the sun \cite{Majorana:2016hop,Panda:2017,LUX,EDELWEISS:2018tde,XENON:2020rca}. For such searches, the consistency requirement on the lifetime of ALPs is greatly reduced. Finally, our results apply only to ALPs, and do not in general extend to other dark matter candidates in the keV--MeV mass range.

\subsection{Acknowledgements}
\begin{acknowledgments}
We are grateful to Edoardo Vitagliano for comments on an earlier version of this paper. The work of RZF was supported by Spanish Ministry of Science and Innovation (PID2020-115845GB-I00/AEI/10.13039/501100011033), the Direcció General de Recerca del Departament d'Empresa i Coneixement (DGR) and by the EC through the program Marie Sklodowska-Curie COFUND (GA 801370)-Beatriu de Pinós. IFAE is partially funded by the CERCA program of the Generalitat de Catalunya. DM and EM are supported by the European Research Council under Grant No. 742104 and by the Swedish Research Council (VR) under grants 2018-03641 and 2019-02337.
\end{acknowledgments}

\bibliography{biblio}

\end{document}